\documentclass[a4paper,12pt]{article}
\usepackage{amssymb}
\usepackage{mathrsfs}
\usepackage{cite}
\usepackage{texdraw}
\usepackage[english]{babel}

\csname @addtoreset\endcsname{equation}{section}

\newcommand{\ybox}[1]{{\bf #1}}

\newcommand{\R}{\mathbb{R}}
\newcommand{\X}{\mathbb{X}}
\newcommand{\Z}{\mathbb{Z}}

\newcommand{\HH}{\mathcal{H}}
\newcommand{\LL}{\mathcal{L}}
\newcommand{\NN}{\mathcal{N}}
\newcommand{\OO}{\mathcal{O}}
\newcommand{\RR}{\mathcal{R}}
\newcommand{\calS}{\mathcal{S}}
\newcommand{\TT}{\mathcal{T}}

\newcommand{\id}{\mathbf{1}}

\newcommand{\dd}{\mathrm{d}}
\newcommand{\tr}{\mathrm{Tr}}

\newcommand{\p}{\partial}

\newcommand{\arr}{\begin{eqnarray*}}
\newcommand{\farr}{\end{eqnarray*}}

\def\al{\alpha}
\def\be{\beta}
\def\ga{\gamma}

\def\ep{\epsilon}
\def\eps{\varepsilon}

\def\la{\lambda}

\def\si{\sigma}
\def\om{\omega}

\def\Ga{\Gamma}

\begin{document}

\begin{titlepage}
\begin{flushright}
IFUM-785-FT
\end{flushright}
\vspace{.3cm}
\begin{center}
\renewcommand{\thefootnote}{\fnsymbol{footnote}}
{\Large \bf Multi giant graviton systems, SUSY breaking and CFT}
\vskip 25mm
{\large \bf {Marco M.~Caldarelli\footnote{marco.caldarelli@mi.infn.it} and
Pedro J.~Silva\footnote{pedro.silva@mi.infn.it}}}\\
\renewcommand{\thefootnote}{\arabic{footnote}}
\setcounter{footnote}{0}
\vskip 10mm
{\small
Dipartimento di Fisica dell'Universit\`a di Milano\\
and\\
INFN, Sezione di Milano,\\
Via Celoria 16, I-20133 Milano.\\
}
\end{center}
\vspace{2cm}
\begin{center}
{\bf Abstract}
\end{center}
{\small In this article, we describe giant gravitons in
  AdS$_5\times\calS^5$ moving along generic trajectories in AdS$_5$.
  The giant graviton dynamics is solved by proving that the D3-brane
  effective action reduces to that of a massive point particle in
  AdS$_5$ and therefore the solutions are in one to one
  correspondence with timelike geodesics of AdS$_5$. All these
  configurations are related via isometries of the background, which
  induce target space symmetries in the world volume theory of the
  D-brane. Hence, all these configurations preserve the same amount of
  supersymmetry as the original giant graviton, i.e. half of the
  maximal supersymmetry.
  Multiparticle configurations of two or more giant gravitons are also
  considered. In particular, a binary system preserving
  one quarter of the supersymmetries is found, providing a non trivial
  time-dependent supersymmetric solution.
  A short study on the dual CFT description of all the above states is
  given, including a derivation of the exact induced isometry map in
  the CFT side of the correspondence.}
\end{titlepage}


\section{Introduction}

The AdS/CFT duality is one of the most celebrated subjects
within string theory (see
\cite{Petersen:1999zh,Aharony:1999ti,D'Hoker:2002aw} for reviews).
After more than six years of continuous studies, this holographic
conjecture has become more and more  robust, passing all the so-called
checks or theoretical experiments that the string theory community
has been able to engineer. Nevertheless, there are many aspects to
understand and uncover in this puzzling correspondence that relates
string theory on anti-de~Sitter (AdS) spaces and conformal field
theories (CFT).

Recently, the study of a new kind of stable D3-brane
configurations on AdS$_5\times\calS^5$ has brought some attention.
These extended solitons are stabilized by a dynamical mechanism,
developing local forces on the brane that cancel its tension,
avoiding the world volume collapse. More precisely, these D-brane
configurations, or {\em giant gravitons}, correspond to D3-branes
travelling on a $\calS^1$ direction wrapping a perpendicular $\calS^3$,
both contained in the $\calS^5$ factor of the metric, while they sit on
the center of AdS$_5$ \cite{McGreevy:2000cw}.
The dynamics of the D-brane effective action allows for two
different stable solutions, one in which the radius of the $\calS^3$ is zero
and the other with a non-zero radius, bounded from above and
proportional to the momentum along the $\calS^1$ direction.
These states preserve half of the supersymmetries
\cite{Grisaru:2000zn} and, from the ten-dimensional point of view,
their geometrical center travels along a null geodesic.
Such solitons are interpreted as supergraviton states
that expand into a sphere following a sort of Myers' effect
\cite{Myers:1999ps}.

Originally, these configurations where thought as the gravitational
manifestation of the stringy exclusion principle
\cite{Maldacena:1998bw}, where the upper bound on the giant graviton
momentum on the $\calS^1$ (due to the fact that it is proportional to the
radius of the $\calS^3$ and therefore has a maximum on $\calS^5$), is
dual to the upper bound found on the conformal weight of a family of
chiral operators (here the bound is easily understood from the
finite rank of the gauge symmetry group)
\cite{McGreevy:2000cw,Balasubramanian:2001nh}.

Nevertheless, giant gravitons have brought new physics, like in the
study of their back reaction, where particular condensates of giant
gravitons result in supersymmetric solutions of type IIB supergravity
called {\em superstars} \cite{superstars}.
Another fascinating characteristic of these giant gravitons is their
ability to regulate potential divergences by enlarging their size while
the energy of the configuration is increased. This behavior,
somehow characteristic in string theory, relates UV and IR regimes and
is certainly telling us that there is a lot to understand on the
reparameterization invariance in the presence of Ramond-Ramond fluxes.
Also, there exists another type of solution, known as {\em dual giant
graviton}, with the same quantum numbers, but whose world volume
grows entirely in AdS$_5$.
In this case, the dual giant graviton size does not have any upper bound
and hence has been related to a different type of chiral operators in a
different representation of the R-symmetry group \cite{dualgg}.

In this article, we shall consider the generalization of giant
gravitons to more complicated configurations in the AdS$_5$ factor of
the ten-dimensional space-time. In section two, we look for
the most general solution of the D3-brane embedded in AdS$_5$ by making
explicit the relation between the dynamics of this type of brane ansatz
and point particles on AdS$_5$. Then, due to the fact that all the
point particles move on geodesics and that all such geodesics are
interrelated by isometry transformations, we obtain all the
possible solutions corresponding to these {\em generalized giant
  gravitons}. In particular, we find supersymmetric configurations
corresponding to giant gravitons rotating on a circular orbit in
AdS$_5$ at a given fixed radius that depends on the angular momenta in
AdS$_5$ and in $\calS^5$. In section three, we study the supersymmetry
properties of the above solutions and find that all of them correspond
to one half supersymmetric configurations. A family of quarter BPS
solutions is proposed by considering multiparticle states of two giant
gravitons. In section four, we comment on the CFT duals and its
supersymmetry properties, to end in section five, with a short summary
and conclusions.

Although the whole article is written for giant gravitons
corresponding to D3-brane configurations, it is trivial to extend the
supergravity discussion and calculations to the M2-brane and M5-brane
giant gravitons of M-theory. Here we have not done so to avoid
unnecessary complications with the notation.

\bigskip
{\em Note added in proof:}
while we were writing this article, the work \cite{Arapoglu:2003ti}
appeared, with some partial overlap with the material here
discussed.


\section{Generalized giant gravitons and their relation with
point particles in AdS$_5$}

In this section we study the generalized giant graviton as an embedded
probe D3-brane in the near horizon geometry of $N$ D3-branes of type IIB
supergravity, i.e. on AdS$_5\times\calS^5$, in such a way that all its
spacelike directions coincide with $\calS^5$ directions, but leaving
the AdS$_5$ motion otherwise free. We compute the reduced action
describing its classical dynamics in AdS$_5$ with frozen internal degrees of
freedom, which turns out to be a massive point particle action.
Therefore, since we are describing point particle dynamics, all possible
solutions correspond to timelike geodesics in AdS$_5$. Note that all
timelike geodesics are interrelated by isometries of the background
and that these transformations correspond to symmetries of the
D-brane action. After proving the above statements, we consider the
particular example of a giant graviton with angular momentum on AdS$_5$.

\subsection{Giant gravitons as point particles}

First, to fix notation and conventions, we choose global coordinates
on AdS$_5\times\calS^5$ such that the AdS$_5$ factor of the metric is
given by
\begin{equation}
  ds^2_{\mathrm{AdS}}=-V\left(r\right)dt^2
  +\frac{dr^2}{V(r)}+r^2\,d\Omega_3^2
\end{equation}
where the lapse function is given by $V(r)=1+r^2/L^2$, and the
transverse three sphere can be parameterized by coordinates
$(\al_1,\al_2,\psi)$, with metric
\begin{equation}
  d\Omega^2_3=d\al_1^2+\sin^2\al_1\left(d\al^2_2+\sin^2\al_2\,d\psi^2\right)\,.
\label{threesphere}\end{equation}
Another useful set of coordinates for the three sphere is
$(\be,\psi_1,\psi_2)$, with metric
\begin{equation}
  d\Omega^2_3=d\be^2+\sin^2\be\,d\psi^2_1+\cos^2\be\,d\psi^2_2
\label{threesphere2}\end{equation}
where now the $SU(2)\times SU(2)$ subgroup of the $SO(4)$ rotation
group is manifest.
For the $\calS^5$ part of the metric, we take
\begin{equation}
  ds^2_{5}=L^2\left(d\theta^2+\cos^2\theta\,d\phi^2
    +\sin^2\theta\,d\omega_3^2\right)
\end{equation}
where
\begin{equation}
  d\omega^2_3=d\chi_1^2+\sin^2\chi_1
  \left(d\chi^2_2+\sin^2\chi_2\,d\chi^2_3\right)
\end{equation}
is the three sphere on which the D3-brane will grow.
Also, we shall use curved indices $M, N, \ldots$ for the full
ten-dimensional metric and indices $\mu, \nu, \ldots$ for the AdS$_5$
part of the metric. A point in AdS$_5\times\calS^5$ is then described by
$X^M=(t,r,\al_1,\al_2,\psi,\theta,\phi,\chi_1,\chi_2,\chi_3)$,
while its projection on AdS$_5$ is denoted
$x^\mu=(t,r,\al_1,\al_2,\psi)$.

The D3-brane low-energy dynamics is described by a Born-Infeld action
with a Chern-Simons coupling, given by
\begin{equation}
  S_\mathrm{D3}=-T_3\int\!\!d^4\si\,\sqrt{-g}+T_3\int a^{[4]}\,,
\end{equation}
where $g$ is the pull back of the space-time metric to the
world volume, i.e.
\begin{equation}
  g_{IJ}=\p_I X^M\p_J X^N G_{MN}\ ,
\end{equation}
$a^{[4]}$ denotes the pull back of the Ramond-Ramond 4-form
potential $A^{[4]}$, and we have used indices $I, J, \ldots$ for the
world volume directions of the D3-brane.
The tension of the brane is $T_3=(8\pi^3g_s\ell_s^4)^{-1}$ where $g_s$
and $\ell_s$ are the coupling constant and length of the string
respectively.

We want to study stable probe D-brane configurations where the
D-brane has expanded on the $\calS^5$ to a three sphere at fixed
$\theta$, while it orbits in the $\phi$ direction, and all
the world volume modes of the D-brane are frozen.
It is convenient to choose an embedding such that the
world volume coordinates $\si^I$ are identified with the
appropriate space-time coordinates,
\begin{eqnarray}
  \si_1=\chi_1,\qquad \si_2=\chi_2,\qquad
  \si_3=\chi_3,\nonumber \\
  x^\mu=x^\mu(\si_0),\qquad \theta=\theta_0,\qquad \phi=\phi(\si_0).
\end{eqnarray}
Then, the pull back of the metric reads
\begin{equation}
  g_{IJ}=\left(
    \begin{array}{cc}
      G_{MN}\dot X^M\dot X^N & 0\\
      0                      & L^2\sin^2\theta(g_\chi)_{ab}
    \end{array}
  \right)\,,
\end{equation}
where $(g_\chi)_{ab}$ is the $\calS^3$ metric corresponding to
$d\omega^2_3$ and the dot stands for the derivative with respect to
$\si_0$. Integrating out the spacelike directions $\si_a$ in the
D-brane action in the above embedding, we obtain the following reduced
action,
\begin{equation}
  S_\mathrm{D3}=-\frac NL\sin^3\theta\int\!\!d\si_0\,
  \sqrt{-G_{MN}\dot X^M\dot X^N}
  +N\sin^4\theta_0\int\!\!\dot\phi\,d\si_0\ .
\end{equation}
Here we have used the relation $L^4=4\pi g_sN\ell_s^4$ characteristic
of D3-brane near horizon backgrounds.
A more convenient and nevertheless equivalent form for this action is
\begin{equation}
  S=\frac12\int\!\!d\si_0\left(\frac1eG_{MN}\dot X^M\dot X^N-m^2e\right)
  +N\sin^4\theta_0\int\dot\phi\,d\si_0\,,
  \label{S}
\end{equation}
where we have defined
\begin{equation}
  m=\frac{N}L\sin^3\theta\,,
\end{equation}
and the {\sl einbein} $e$ plays the role of a Lagrange
multiplier\footnote{To check the equivalence between the two
actions, one can simply use the variation of $S$ to find
$e$,
\begin{equation}
  e=\frac1m\sqrt{-G_{MN}\dot X^M\dot X^N}
\end{equation}
and then substitute its value in $S$ to recover $S_\mathrm{D3}$.}.
Separating the AdS$_5$ component of the motion
\begin{equation}
  G_{MN}\dot X^M\dot X^N=g_{\mu\nu}\dot x^\mu\dot x^\nu
  +L^2\cos^2\theta\ \dot\phi^2\,,
\end{equation}
the action reads
\begin{equation}
  S=\frac12\int\!\!d\si_0\left(\frac1eg_{\mu\nu}\dot x^\mu\dot x^\nu
    -m^2e+\frac{L^2}e\cos^2\theta_0\,\dot\phi^2
    +2N\sin^4\theta_0\,\dot\phi\right)\,.
\end{equation}
We see that the coordinate $\phi$ is cyclic, hence its conjugate momentum
\begin{equation}
  p_\phi=\frac{L^2}e\dot\phi\cos^2\theta+N\sin^4\theta
\label{pphi}\end{equation}
is conserved. It is therefore useful to define the Routh function
\begin{equation}
  \RR(x^\mu, \dot x^\mu, p_\phi, \theta_0) = \dot\phi p_\phi-\LL.
\end{equation}
After some algebra, we obtain
\begin{equation}
  \RR(x^\mu, \dot x^\mu, p_\phi, \theta_0) =
  -\frac1{2e}g_{\mu\nu}\dot x^\mu\dot x^\nu
  +\frac e{2L^2}\left[p_\phi^2+\tan^2\theta_0
    \left(p_\phi-N\sin^2\theta_0\right)^2\right].
\end{equation}
Since the time derivative of $\theta$ does not appear in the
routhian, the equations of motion $\p\RR/\p\theta=0$ for this coordinate
yield the constraint
\begin{equation}
  \tan\theta\left(p_\phi-N\sin^2\theta\right)
  \left(p_\phi-N\sin^2\theta\left(1+2\cos^2\theta\right)\right)=0\,.
\end{equation}
There are two stable minima for the above potential, $\theta_0=0$
corresponding to a collapsed D3-brane and
\begin{equation}
  \sin^2\theta_0=\frac{p_\phi}N,
  \label{radiuss5}
\end{equation}
representing an expanded D-brane, or giant graviton, of radius
$\sin\theta_0$. Finally, there is an unstable maximum of the
potential between these two. In what follows, we will restrict to
the giant graviton case. We can then substitute (\ref{radiuss5})
in the Routh function, and obtain
\begin{equation}
  \RR=-\frac12\left(
  \frac1{e}g_{\mu\nu}\dot x^\mu\dot x^\nu
  -e\frac{p_\phi^2}{L^2}\right).
\end{equation}
Now, this is just (minus) the lagrangian for a particle of mass
\begin{equation}
M=\frac{p_\phi}L
\label{mass}
\end{equation}
moving in AdS$_5$.

In other words, the general solution $x^\mu(\si_0)$ of the
dynamical problem of the above D-brane is described by the
dynamics of a point particle with mass $M$ given in eq. (\ref{mass}),
which is solved by timelike geodesics\footnote{In
  \cite{Bozhilov:2002sj}, the conditions under which probe brane
  dynamics reduces to a particle-like one were analysed.}.
Therefore, {\em from the AdS$_5$ point of view, giant gravitons are just
massive point-like particles propagating along timelike geodesics}.

\subsection{Generalized giant graviton solutions}
\label{ggg}

To obtain the actual form of the generalized giant
graviton solutions, we use the static gauge setting $\dot x^0=1$.
The conjugate momenta to the remaining dynamical variables are
\begin{equation}
  p_i=\frac1eg_{ij}\dot x^j,\qquad
\label{pi}\end{equation}
where $i, j, \ldots$ are the spatial directions on AdS$_5$,
and the hamiltonian reads
\begin{equation}
  \HH=\dot x^ip_i+\RR=\frac{V(r)}{2e}+\frac e2\left(g^{ij}p_ip_j+M^2\right)\,.
\end{equation}
The equation for $e$ can be solved, yielding
\begin{equation}
  e=\sqrt\frac{V(r)}{g^{ij}p_ip_j+M^2}\,;
\end{equation}
plugging this back in the hamiltonian we obtain
\begin{equation}
  \HH=\sqrt{V(r)}\sqrt{g^{ij}p_ip_j+M^2}\,.
\end{equation}
To make manifest the full symmetry of the problem, we use the coordinates
$(\be,\psi_1,\psi_2)$ with metric (\ref{threesphere2}) for $\calS^3$.
The explicit form of the hamiltonian is then
\begin{equation}
  \HH=\sqrt{V(r)}\sqrt{
    V(r)p_r^2+\frac1{r^2}\left(
      p_\beta^2+\frac{p_{\psi_1}^2}{\sin^2\beta}
      +\frac{p_{\psi_2}^2}{\cos^2\beta}
    \right)+M^2}\,.
\end{equation}
Note that there is rotation symmetry in $\psi_1$ and $\psi_2$ and hence
$p_{\psi_1}$ and $p_{\psi_2}$ are conserved.
One can then easily check that $J$, defined by
\begin{equation}
  J^2=p_\beta^2+\frac{p_{\psi_1}^2}{\sin^2\beta}
      +\frac{p_{\psi_2}^2}{\cos^2\beta}\,,
\end{equation}
is also first integral of the hamiltonian\footnote{This
  further conserved quantity does not descend from a spacetime
  isometry, but rather from a St\"ackel-Killing tensor, coinciding
  with the Casimir invariant of any of the $SU(2)$ subgroups of
  $SO(4)$. As a consequence, $(J,p_{\psi_1},p_{\psi_2})$ completely
  determine the angular motion. See \cite{Gibbons:1999uv} for further
  details.}, and allows to decouple completely the angular motion from
the radial one. The projection on $\calS^3$ of the motion will always
be a constant point or describe a movement on a great circle of the
three-sphere.
Finally, the radial motion of the brane is determined by the hamiltonian
\begin{equation}
  \HH=\sqrt{V(r)}\sqrt{
    V(r)p_r^2+\frac{J^2}{r^2}+M^2}.
\end{equation}
This hamiltonian is a constant of motion, and its value is the
energy $E$ of the giant graviton. This is enough to solve the
one-dimensional radial motion; we have
\begin{equation}
  p_r=\frac{E\dot r}{V^2(r)}\ ,
\end{equation}
hence
\begin{equation}
  E^2\dot r^2=V^2(r)\left[E^2-\left(M^2+\frac{J^2}{r^2}\right)V(r)\right]
  \label{eqradial}
\end{equation}
which can be readily integrated to obtain the most general
solution (see for example \cite{Dorn:2003ct} for the explicit solutions).

As an example, which will prove interesting in the subsequent
sections, let us consider solutions with constant radius $r_0$. Then
$p_r=0$, and the equation of motion $\p\HH/\p r=0$ yields
\begin{equation}
  r_0=L\sqrt\frac{J}{p_\phi}. \label{radius}
\end{equation}
The giant graviton rotates at constant velocity on a great circle
of radius $r_0$ of the transverse $\calS^3$ in AdS$_5$, with angular
momentum $J$ with projections $p_{\psi_1}$ and $p_{\psi_2}$ on the
$\psi_1$ and $\psi_2$ axes. The total energy of this configuration is
then obtained by substituting $r$ by its value in the hamiltonian
$\HH$, to obtain
\begin{equation}
  E=\frac{J+p_\phi}L.
\label{energy}\end{equation}
The energy is linear in the conserved quantities, and this is
reminiscent of a BPS bound. In fact, as we will show in the next
section, all these solutions preserve one half of the
supersymmetries.

It is interesting to see the dynamics of the D3-brane from the
ten-di\-mensional point of view. The total hamiltonian
obtained by Legendre-transforming all the canonical
variables directly in the initial action (\ref{S}) reads
\begin{equation}
  \HH=\frac e2\left[
    g^{\mu\nu}p_\mu p_\nu+\frac1{L^2\cos^2\theta}
    \left(p_\phi-N\sin^4\theta\right)^2+\frac{N^2}{L^2}\sin^6\theta
    \right].
\end{equation}
This hamiltonian is linear in the Lagrange multiplier $e$; it is a
pure constraint, imposing the vanishing of the hamiltonian.
After substituting for the momenta, using equations (\ref{pphi}) and
(\ref{pi}), it translates into
\begin{equation}
  G_{MN}\dot X^M\dot X^N=-\frac{e^2N^2}{L^2}\sin^6\theta<0\,.
\end{equation}
This means that each world volume element of the D3-brane follows
a timelike trajectory. However, the geometrical center of the brane,
which is located at $\theta=0$, follows a null trajectory in the
full ten-dimensional space-time.

\subsection{Anti-de~Sitter isometries and geodesics}

In the above subsection, we have shown that giant gravitons follow
timelike geodesics in AdS$_5$. Also, it is important to note that in
a homogeneous spacetime, timelike geodesics can be mapped one into
another by means of isometry transformations. This can be easily
proved using the following standard proposition (see for example
\cite{oneill})

\bigskip
\noindent{\bf Proposition 1:}
{\sl Let $e_1,\ldots,e_5$ and $f_1,\ldots,f_5$ be tangent frames on
  AdS$_5$ at points $p$ and $q$, respectively. Then there is a unique
  isometry $\phi:\R^4_2\rightarrow\R^4_2$ carrying AdS$_5$
  isometrically to itself, with $\phi(p)=q$ and $\phi_*(e_i)=f_i$ for
  $i=1\ldots5$.}
\bigskip

Then, using the fact that any geodesic is uniquely determined by a
point $p$ and a unit timelike tangent vector, we find,

\bigskip
\noindent{\bf Theorem 1:} {\sl Every timelike geodesic of AdS$_5$
can be mapped on any other timelike geodesic by an isometry. }
\bigskip

As a consequence, an alternative way to construct the generic giant
graviton solution of the previous subsection is to start with any
given particular solution and transform it by acting with an AdS$_5$
isometry.

We would like to stress that the above isometries of the
background translate into target space symmetries of the world volume
theory of the D-brane, and therefore map solutions into solutions. 
In fact, we will argue in the next section that this target space
symmetry leaves the full supersymmetric action invariant, and hence
all these solutions preserve the same amount of supersymmetry.

To make our discussion more specific, we introduce in this
subsection a mathematical formalism to handle the isometries in a
pleasant form that will prove useful in forthcoming sections.

AdS$_5$ is a homogeneous five dimensional manifold that can be
embedded as the hyperboloid
\begin{equation}
-X_0^2+X_1^2+X_2^2+X_3^2+X_4^2-X_5^2=-L^2
\label{AdS}\end{equation}
in flat six-dimensional space-time with signature $+2$,
\begin{equation}
\dd s^2=-\dd X_0^2+\dd X_1^2+\dd X_2^2+\dd X_3^2+\dd X_4^2-\dd X_5^2\,.
\end{equation}
The hyperboloid is manifestly  invariant under the $SO(4,2)$
isometry group of the embedding space, and can be parameterized with
coordinates $(\tau,\rho,\al_1,\al_2,\psi)$ such that
\begin{eqnarray}
X_0 = L\cosh\rho\cos\tau, & X_5= L\cosh\rho\sin\tau,&\\
X_1= L\sinh\rho\cos\al_1, & X_2= L\sinh\rho\sin\al_1\cos\al_2,&\nonumber\\
X_3= L\sinh\rho\sin\al_1\sin\al_2\cos\psi,
&X_4= L\sinh\rho\sin\al_1\sin\al_2\sin\psi.&\nonumber
\end{eqnarray}
In these coordinates, the induced AdS$_5$ metric reads\footnote{In
  this subsection, we use here the new dimensionless coordinates
  $\rho$ and $\tau$, defined as $L\,\dd\rho=V^{-1}(r)\,\dd r$ and
  $\tau=t/L$.
}
\begin{equation}
  \dd s^2=L^2\left(-\cosh^2\rho\ \dd\tau^2+\dd\rho^2
  +\sinh^2\rho\ \dd\Omega_3^2\right)
\end{equation}
where $\dd\Omega_3^2$ is the three-sphere metric given in equation
(\ref{threesphere}).
Here $\tau\in[0,2\pi)$, $\rho\in\mathbb{R}_+$, $\al_i\in[0,\pi)$,
$\psi\in[0,2\pi)$. To avoid closed timelike curves, we take the
universal cover of the hyperboloid, by simply allowing the time
coordinate $\tau$ to range on the whole real axis.

The following propositions \cite{oneill} give a geometrical picture
of the geodesics on this hyperboloid,

\bigskip
\noindent{\bf Proposition 2:}
{\sl Let $\ga$ be a nonconstant geodesic of AdS$_5\subset\R^4_2$. If
  $\ga$ is spacelike it is a parameterization of one branch of a
  hyperbola in $\R^4_2$. If $\ga$ is null, it is a straight line, that
  is, a geodesic of $\R^4_2$. If $\ga$ is timelike it is a periodic
  parameterization of an ellipse in $\R^4_2$.
}

\bigskip
\noindent{\bf Proposition 3:} {\sl The geodesics of
AdS$_5\subset\R^4_2$ are the curves sliced from AdS$_5$ by planes
$\Pi$ through the origin of $\R^4_2$.}
\bigskip

To give an explicit example of solution construction by isometries, we
shall work out the case of the giant graviton moving at constant
radius on the equatorial plane $\al_1=\pi/2$, $\al_2=\pi/2$ found in
the previous subsection. From the
embedding space point of view, we restrict to the subspace $\{X_1=0,
X_2=0\}$. The points in the equatorial plane form the AdS$_3$ hyperboloid
\begin{equation}
  \label{eq:AdS3}
  -X_0^2+X_3^2+X_4^2-X_4^2=-L^2\,.
\end{equation}
Hence, a point $x\in \mathrm{AdS}_3$ can be parameterized by the
$SL(2,\mathbb{R})$ matrix \cite{Matschull:1998rv}
\begin{equation}
  \X=\frac1L\left(
    \begin{array}{cc}
      X_0+X_3  & X_5+X_4\\
      -X_5+X_4 & X_0-X_3
    \end{array}
  \right)\,,
\end{equation}
because $\det\X=1$ is equivalent to condition (\ref{eq:AdS3}). In terms
of the coordinates, the explicit matrix is
\begin{equation}
  \X(\tau,\rho,\psi)=\left(
    \begin{array}{cc}
      \cosh\rho\cos\tau+\sinh\rho\cos\psi &
      \cosh\rho\sin\tau+\sinh\rho\sin\psi \\
      -\cosh\rho\sin\tau+\sinh\rho\sin\psi &
      \cosh\rho\cos\tau-\sinh\rho\cos\psi
    \end{array}
  \right)\,.
\end{equation}
The residual isometry group is $SO(2,2)\subset SO(4,2)$. Using the
isomorphism $SO(2,2)\cong SL(2,\R)\times SL(2,\R)/\Z_2$, its
action on the point $\X$ on the equatorial plane is given by
\begin{equation}
  (\rho_L,\rho_R): \X\mapsto\rho_L\X\rho_R
\end{equation}
with $\rho_L, \rho_R\in SL(2,\R)$, and the $\Z_2$ quotient is obtained by the identification of $(\rho_L,\rho_R)$ with $(-\rho_L,-\rho_R)$.
A basis of the $sl(2,\R)$ algebra is given by the matrices
\begin{equation}
  {\bf 1}=\left(\begin{array}{cc} 1 & 0 \\ 0 & 1 \end{array}\right),
  \ga_0=\left(\begin{array}{cc} 0 & 1 \\ -1 &  0 \end{array}\right),
  \ga_1=\left(\begin{array}{cc} 0 & 1 \\  1 &  0 \end{array}\right),
  \ga_2=\left(\begin{array}{cc} 1 & 0 \\  0 & -1 \end{array}\right)
\end{equation}
\[
\ga_a\ga_b=\eta_{ab}\id-\eps_{ab}{}^c\ga_c\,,\qquad
\eta_{ab}=\mathrm{diag}(-1,1,1)\,,\qquad
\eps^{012}=1\,,
\]
which generate the following elements of the group,
\begin{equation}
  {\cal R}(\al)=e^{\al\ga_0}\,,\qquad
  {\cal S}(\al)=e^{\al\ga_1}\,,\qquad
  {\cal T}(\al)=e^{\al\ga_2}\,.
\end{equation}
A useful relation is
$\calS(\al)=\RR\left(-\pi/4\right)\TT(\al)\RR\left(\pi/4\right)$.
Now $(\tau,\rho,\psi)$ are essentially the Euler angles of $SL(2,\R)$,
\begin{equation}
  \X=e^{\frac12(\tau-\psi)\ga_0}e^{\rho\ga_1}e^{\frac12(\tau+\psi)\ga_0}
\end{equation}
and any point of the equatorial plane may be written
\begin{equation}
  \X(\tau,\rho,\si)=\RR\left(\frac{\tau-\psi}2\right)\,\TT(\rho)\
  \RR\left(\frac{\tau+\psi}2\right)\,.
  \label{X2}
\end{equation}

Let us start with the original giant graviton sitting in the center of
AdS$_5$ found in ref.~\cite{McGreevy:2000cw}. This
giant graviton is described by the timelike geodesic $\rho=0$ (a
particle at rest in the center of AdS$_5$). We can take for convenience
$\psi=0$, then its trajectory is parameterized by
\begin{equation}
  \X(\la)=\RR(\la)\,,\qquad \la\in\R\,.
\end{equation}
Next consider the isometry $(\TT(\rho_0),\id)$. This transformation
maps the old geodesic into the new geodesic
\begin{equation}
  \X'(\la)=\TT(\rho_0)\RR(\la).  
\end{equation}
From eqn.~(\ref{X2}), we can rewrite the trajectory as
$\X'(\la)=\X(\la,\rho_0,\la)$
and hence $\{\psi=\tau,\rho=\rho_0\}$,
i.e. the particle moves on a circle of radius $\rho=\rho_0$ at constant
velocity $\p_t\psi=1/L$.
In this particular case, from the hamiltonian analysis, we get that
$J=p_\psi=Er^2/LV(r)$. Then, using equation
(\ref{energy}), we find the expected value for the radius, in
agreement with (\ref{radius}).

From the point of view of the embedding space, the isometry transformation
$(\TT(\rho_0),\id)$ is the composition of two hyperbolic rotations in the
$(0,3)$ and $(4,5)$ planes, both with angle $\rho_0$:
\begin{equation}
  (\TT(\rho_0),\id)=R_{(0,3)}(\rho_0)\circ R_{(4,5)}(\rho_0)\ .
\end{equation}

Note that the two planes are orthogonal and the rotations commute. Moreover,
the $X_1$ and $X_2$ coordinates are left untouched by this transformation,
hence
\begin{equation}
  X_1'=X_1,\qquad X_2'=X_2\,,
\end{equation}
and the corresponding infinitesimal generator is given by
\begin{equation}
  T=X_0\p_3+X_3\p_0+X_4\p_5+X_5\p_4.
\end{equation}
In particular, the above infinitesimal transformation will be of
relevance in the study of the dual operators in the CFT.


\section{One half and less supersymmetric states}

In this section, we study the supersymmetry properties of
generalized giant graviton configurations in AdS$_5\times\calS^5$.
Then, we consider particular examples starting with single particle
states, to end with multiparticle states, having in mind the idea of
breaking some additional supersymmetry.

Bosonic D-brane and M-brane configurations living in bosonic
supersymmetric backgrounds are supersymmetric if the background
Killing spinor $\ep$ satisfies the $\kappa$-symmetry constraint
\begin{equation}
  \left(1+\Gamma\right)\epsilon=0\,,
\label{k}\end{equation}
where $\Gamma$ is the relevant $\kappa$-symmetry matrix \cite{M2,M5,D}.
The above equation dictates the form and number of real independent
parameters that produce supersymmetric transformations on the world
volume theory of the brane.

The background Killing spinor can always be written as
$\Pi(x)\ep_0$, with $\Pi(x)$ a general space-time dependent
matrix. Its rank is constant and equal to the number of supersymmetry
generators which leave invariant the spacetime fields; the rank of
the matrix $M(x)\equiv(1-\Gamma(x))\Pi(x)$ counts the number of
surviving world volume supersymmetries.

Note that isometry transformations of the background translate into
target space symmetries of the world volume theory, therefore mapping
solutions into physically distinct solutions.
The $\kappa$-symmetry condition (\ref{k}) is covariant under general
diffeomorphisms, in particular it is left invariant in form under
isometries.
Also, $M(x)$ transforms as a scalar field under general diffeomorphisms
and if we act with an isometry its rank remains unchanged. Therefore,
{\em the transformed brane solutions preserve the same amount of
  supersymmetries.}

The general giant graviton solutions found in the previous
section are related via isometries to the giant graviton solution
sitting in the center of AdS$_5$, which is known to be one half
supersymmetric \cite{Grisaru:2000zn}. Consequently, applying the
previous argument, we deduce that {\em all generalized giant gravitons
  are half BPS states}.

Nevertheless, the particular form of the surviving supersymmetry
generator depends on the specific form of the solution. Accordingly,
to obtain configurations with less preserved
supersymmetry, we can consider multiparticle giant graviton states,
such that the relative motion breaks some further fraction of supersymmetry.
Observe that this strategy to construct fractional BPS states using
isometries, is not peculiar to giant gravitons but can be applied to
any multibrane configuration.

In what follows, we will illustrate the above arguments with specific
cases in order to provide explicit examples.

\subsection{Giant graviton with angular momentum on AdS$_5$}

Let us consider the solution describing a giant graviton located at
constant radial position $r_0$ given in eqn.~(\ref{radius}) and
rotating on one of the great circles of the $\calS^3$, with constant
angular velocity $\dot\psi=1/L$. The corresponding embedding is
\begin{eqnarray}
  \si_0\equiv\tau=t,\quad\si_1=\chi_1,\quad\si_2=\chi_2,\quad
  \si_3=\chi_3,\nonumber \\
  r= r_0,\quad \psi=\tau/L,\quad \theta=\theta_0,\quad \phi=\tau/ L,
  \nonumber \\
  \alpha_1=\alpha_2=\pi/2.
\label{embedding}
\end{eqnarray}
where the values of $(r,\theta)$ can be parameterized by the
corresponding angular momenta on  AdS$_5$ and $\calS^5$ as
\begin{equation}
p_\phi=N\sin^2\theta_0\,,\qquad
\frac{J}{p_\phi}=\left(\frac{r_0}{L}\right)^2.
\end{equation}
To write the $\kappa$-symmetry constraint we need to set the following
definitions and conventions: we label ten-dimensional
tangent space indices by $A, B, \ldots$, such that the vielbein is
written as $e_M^A$; a particular value of a Lorentz index is underlined, e.g.
$\underline\psi$, while curved space-time indices are left
unadorned. We use a real representation of the ten-dimensional
$\Gamma$-matrices $\Gamma_A$ such that
$\left\{\Gamma_A,\Gamma_B\right\}=2\eta_{AB}$, where
$\eta_{AB}=\mathrm{diag}(-1,1,\ldots,1)$ and $\Gamma_{A_1A_2\ldots
  A_n}$ is a completely antisymmetrized on indices
$(A_1,A_2,\ldots,A_n)$ with weight one.
Finally, we combine the two real Majorana-Weyl Killing spinors
of type IIB supergravity into a single complex Majorana-Weyl spinor
$\epsilon$, that satisfies the following Killing condition
\begin{equation}
\left(D_M - {i\over 4}(\ga_5+\ga)\Ga_M\right) \ep = 0\,,
\end{equation}
where $D_M$ is the ten-dimensional covariant derivative,
$\gamma_5=\Gamma^{\underline{\theta\phi\chi_1\chi_2\chi_3}}$ and
$\gamma=\Gamma^{\underline{t\rho\alpha_1\alpha_2\psi}}$.
The solution to this equation is\footnote{See for example
  \cite{Grisaru:2000zn}. Due to conventions, some signs are different.}
\begin{eqnarray}
\ep=\left[
e^{{i\over 2}\theta\ga_5\Ga^{\underline\theta}}
e^{{i\over 2}\phi\ga_5\Ga^{\underline\phi}}
e^{-{i\over 2}\chi_1\Ga^{\underline{\chi_1\theta}}}
e^{-{i\over 2}\chi_2\Ga^{\underline{\chi_2\chi_1}}}
e^{-{i\over2}\chi_3\Ga^{\underline{\chi_3\chi_2}}}
\times \right. \nonumber \\
\left. e^{{i\over 2}\al\Ga^{\underline{r}}\ga}
  e^{-{it\over 2L}\Ga^{\underline{t}}\ga}
    e^{-{1\over 2}\al_1\Ga^{\underline{\al_1r}}}
    e^{-{1\over2}\al_2\Ga^{\underline{\al_2\al_1}}}
    e^{-{1\over 2}\psi\Ga^{\underline{\psi\al_2}}}\right]
  \ep_0\,, \label{killing}
\end{eqnarray}
where $\sinh\al=r/L$ and $\ep_0$ is a general constant complex spinor.

With the above definitions, in the particular case of a D3-brane, the
$\kappa$-symmetry constraint (\ref{k}) becomes
\begin{equation}
  \left(1-i\Gamma\right) \ep = 0\ ,
\end{equation}
where
\begin{equation}
\Gamma={1\over 4!}\eps^{IJKL}\Gamma_{IJKL}\ ,\quad
\Gamma_I=\partial_IX^Me_M^A\Gamma_A\ .
\end{equation}
Using the embedding (\ref{embedding}), we get the equation
\begin{equation}
  \left[\sqrt{1+\left(r/L\right)^2}
    \Gamma_{\underline{t\chi_1\chi_2\chi_3}}+
    \left(r/L\right)\Gamma_{\underline{\psi\chi_2\chi_3}}
    +\cos\theta\Gamma_{\underline{\phi\chi_2\chi_3}}
    -i \sin\theta\right]\epsilon=0\,.
\end{equation}
This expression can be simplified using the following relations
\begin{eqnarray}
  \Ga_{\underline{t\chi_1\chi_2\chi_3}}=
  -\ga^5\Ga^{\underline\theta}\Ga_{\underline{t\phi}}\,,\quad
  \Gamma_{\underline{\psi\chi_2\chi_3}}=
  -\ga^5\Ga^{\underline\phi}\Ga_{\underline{\psi\phi}}\,,\quad
  \Gamma_{\underline{\phi\chi_2\chi_3}}=
  \ga^5\Ga^{\underline\theta}\,,
\nonumber\\
  \cos\theta-i\sin\theta\ga\Ga^{\underline\phi}=
  e^{-i\theta\ga\Ga^{\underline\phi}}\,,\quad 
  \cos\al\Ga_{\underline{t\phi}}-\sin\al\Ga_{\underline{\psi\phi}}=
  \Ga_{\underline{t\phi}}e^{-\al\Ga_{\underline{t\psi}}}\,,
\end{eqnarray}
to obtain
\begin{equation}
\left[
\Ga^{\underline{t\phi}}e^{\alpha\Ga^{\underline{t\psi}}}
+e^{-i\theta\ga^5\Ga^{\underline \theta}}\right]\ep=0\,.
\end{equation}
Next, we pull the above operator through the space-time dependent
part of $\epsilon$ (see eqn. \ref{killing}), using the $\Gamma$-matrix
algebra. After a long but straightforward calculation, we arrive to
\begin{equation}
\left[\left(\sinh^2\al + \sinh\al
    \cosh\al e^{-({it\over L}\Ga^{\underline t}\ga
      -\psi\Ga^{\underline{\psi\al_2}})}\Ga^{\underline{t\psi}}\right)
  \left(1-i\Ga^{\underline{r\al_1}}\right)+\left(1+\Ga^{\underline{t\phi}}\right)\right]\ep_0=0\,.
\end{equation}
At first sight, it may appear that only one quarter of the
supersymmetries will survive, demanding that both projectors
\begin{eqnarray}
P_{t\phi}={1\over 2}(1+\Ga^{\underline{t\phi}})\qquad \mathrm{and}\qquad
P_{r\al_1}={1\over 2}(1-i\Ga^{\underline{r\al_1}})
\end{eqnarray}
annihilate the spinor $\epsilon_0$. However, the general solution to this
equation is found by decomposing the spinor $\ep_0$ into four
independent components defined by the above projectors, i.e.
\begin{equation}
  \ep_0=\ep^{++}+\ep^{+-}+\ep^{-+}+\ep^{--}\,,
\label{deco}\end{equation}
where
\begin{equation}
P_{t\phi}\ep^{+\pm}=\ep^{+\pm}\,, \quad
P_{t\phi}\ep^{-\pm}=0\,,\quad
P_{r\al_1}\ep^{\pm+}=\ep^{\pm+}\,, \quad
P_{r\al_1}\ep^{\pm-}=0\,.
\end{equation}
Using the relation $\psi= t/L$ and more $\Gamma$-matrix algebra, we get
\begin{equation}
  \cosh\al\left(\cosh\al+\sinh\al\Gamma^{\underline{t\psi}}\right)\ep^{+-}
  +\sinh\al\left(\sinh\al+\cosh\al\Gamma^{\underline{t\psi}}\right)\ep^{--}+
  \ep^{++}=0\,.
\label{kfinal}
\end{equation}
Projecting this equation with $P_{\rho\al_1}$, we obtain
$\epsilon^{++}=0$, and hence the supersymmetry condition reduces to
\begin{equation}
  \cosh\al\left(\cosh\al+\sinh\al\Gamma^{\underline{t\psi}}\right)\ep^{+-}+
  \sinh\al\left(\sinh\al+\cosh\al\Gamma^{\underline{t\psi}}\right)\ep^{--}=0\,.
\end{equation}
From this equation we can read off the final conditions on the Killing
spinor, i.e.
\begin{equation}
  \ep^{++}=0 \qquad \mathrm{and} \qquad
  \ep^{+-}=\tanh\al\Ga^{\underline{\psi t}}\ep^{--}\ .
\label{Susy}\end{equation}
Therefore, $\ep^{-+}$ and $\ep^{--}$ are unconstrained and parameterize, as
expected, a total of $8+8=16$ independent supersymmetries: the
solution preserves exactly half of the supersymmetries.

\subsection{One quarter BPS states as binary systems}
\label{quarter}

We have found that any giant graviton moving in AdS$_5$ behaves like a
half BPS particle.
Therefore, no new breaking of supersymmetry occurs by allowing angular
momenta in AdS$_5$ and/or radial time dependence. To obtain smaller
fractions of supersymmetry, what we can certainly do is to consider
the case of two or more giant gravitons travelling along different
geodesics on AdS$_5$.
To make the calculation tractable, we shall require that the
ten-dimension distance between the giant gravitons remains always
larger than the string length, in such a way that the abelian
probe-brane approximation holds.
In general, we expect that no supersymmetry would survive. Nevertheless,
fine tuning the configurations may lead to one quarter or smaller
fractions of residual supersymmetry.

For example, let us consider a configuration of two giant gravitons,
such that the first one sits at $r=0$ while the second orbits at
$r=r_0$. In this binary system, there is a net angular momentum
between the two giant gravitons that cannot be removed by any isometry.

The supersymmetry of such a configuration is easily checked in the
abelian regime, where we can neglect the interactions between the two probe
D-branes. Then, the total hamiltonian is just the sum of each hamiltonian,
and two different supersymmetry constraints have to be imposed on the
Killing spinor $\ep$. Following previous works \cite{Grisaru:2000zn},
it is not difficult to see that the relevant projector for the first
giant graviton located at $r=0$ is
\begin{equation}
P_{t\phi}\ep_0=0\,.
\end{equation}
Using the decomposition (\ref{deco}) of $\ep_0$ by projectors
$P_{t\phi}$ and $P_{r\al_1}$, we get that
\begin{equation}
\ep^{+\pm}=0\,,
\label{susette}\end{equation}
while for the second giant graviton, we found in eqn. (\ref{Susy})
\begin{equation}
  \ep^{++}=0 \qquad \textrm{and} \qquad
  \ep^{+-}=\tanh\al\,\Ga^{\underline{\psi t}}\ep^{--}\,.
\end{equation}
Since the two giant gravitons share the projector $P_{t\phi}$, we see
immediately that to solve both conditions simultaneously we must impose
$\ep^{--}=0$ in addition to (\ref{susette}).
Therefore only $\ep^{-+}$ is unconstrained, meaning that only eight real
supersymmetries survive, or simply that one quarter of the original
supersymmetry is preserved.

It would be interesting to proceed further and use this technique to
obtain more exotic configurations, with smaller fractions of
supersymmetry. For example, one could try to construct solutions
preserving one eighth of the supersymmetries by adding to the above
binary system a third D-brane projecting out another half of the
residual supersymmetries. As we will see shortly, such states may have
interesting CFT duals, but go beyond the scope of the present work and
we leave them for future investigations.


\section{Comments on the dual CFT picture}

In this section we review some known facts about the AdS/CFT
duality in order to study the generalized giant graviton from the CFT
side. Then we make a few comments on
the meaning of an isometry of AdS$_5$ from the point of view of the
CFT, and how this transformation acts on chiral primary fields and on
the supercharges.
Once this is clarified, we describe how the candidate for the giant
graviton dual operator transforms. As a last point, we take the
obvious and naive candidate operator describing the multiparticle
state of two giant gravitons, and argue that it preserves one quarter
of the total supersymmetry, as it should be from the supergravity picture.

\subsection{Giant gravitons and CFT operators}

The remarkable conjecture of Maldacena, relating type IIB
superstring theory on AdS$_5\times\calS^5$ with $\NN=4$ super Yang-Mills
(SYM) in four dimensions has been successfully put under several tests
until now. Both theories have different realizations of the
same superconformal group. In AdS$_5\times\calS^5$ we have the isometry
group $SO(2,4)\times SO(6)$ or better its covering group
$SU(2,2)\times SU(4)$ (since spinors are also involved on this
background) and there are thirty-two real supercharges that enhance the
invariance group to the supergroup $SU(2,2|4)$. On the field theory
side, the $SU(2,2)$ part is realized as the conformal group of
flat four-dimensional Minkowski space-time, while the $SU(4)$ part
corresponds to the R-symmetry group. Although, at first sight, there
are only sixteen real supercharges $Q$, the extension to the superconformal 
group provides the necessary sixteen extra real supercharges $S$, to reach
the grand total of thirty-two real supercharges.

The checks of this conjecture are mostly restricted to the
strong coupling limit of the 't~Hooft coupling constant, in the
large $N$ approximation of the SYM theory, corresponding to the
supergravity regime of superstring theory. In this limit, the
analysis of Kaluza-Klein excitations due to compactification
on $\calS^5$ leads to several families of field modes with
well-defined transformation properties under the $SU(2,2|4)$ group. At
this point, a study of superconformal representations is needed, since
the conjecture translates into a series of predictions concerning
the spectrum of SYM operators. In particular, {\it short}
representations are specially useful due to the fact that some of
their properties are protected from quantum corrections.
In fact, chiral fields (fields belonging to these representations) in SYM
theory correspond to Kaluza-Klein harmonics on the gravity side.

Primary fields are defined as fields annihilated by all
supercharge operators $S$ and all generators of special conformal
transformations $K$ at the origin. Chiral primary fields are
additionally annihilated by some of the $Q$. For example, we construct
the half BPS family by considering symmetric traceless combination of
the scalar fields $\Phi^I$ of $\NN=4$ SYM of the form $\OO^{I_1\ldots
  I_n}=\tr\left(\Phi^{(I_1}\cdots\Phi^{I_n)}\right)$. These operators
have protected scaling dimension $\Delta$, coinciding with 
their R-symmetry charge; $\Phi^I$ is in the {\bf 6} representation of the
R-symmetry group $SU(4)$ and therefore $\OO^{I_1\ldots I_n}$ has
weight $(0,n,0)$, which matches precisely one of the unitarity bounds for
short representations of the superconformal group\footnote{See
  \cite{Ferrara:1999ed} for a short review.}. The full chiral
multiplet is generated by the repeated
action of the operators $Q$ and $P$ on the chiral primary. All the
multiplet is annihilated by some of the $Q$ and, due to the structure
of the superconformal algebra $[Q,K]\sim S$, half of the $S$ also give
zero on all the states of the multiplet, recovering in this way the
notion of sixteen conserved real supersymmetries, i.e. eight generated
by the $Q$ and eight by the $S$.

Normally, single trace operators in the CFT side are related to
single particle states in the gravity side since, in the large $N$
limit, single trace operators form an orthogonal set. Nevertheless,
this is only correct if the R-symmetry charge of the single trace
operators is not comparable with $N$. If this is not the case,
the orthogonality property is lost, and we have to use a
different type of operators to describe the corresponding dual
single particle states. Giant gravitons are among this type of
particles with very high R-charge. Therefore, they are not
expected to be described by single trace operators. In
\cite{Balasubramanian:2001nh}, subdeterminant operators of the SYM
theory were proposed as the main candidates to describe the original
giant graviton sitting at the center of AdS$_5$,
\begin{equation}
\det{}_n(\Phi)={1\over n!}\varepsilon_{i_1\ldots i_{N-n}j_1\ldots
j_n}\varepsilon^{i_1\ldots i_{N-n}k_1\ldots k_n}\Phi^{j_1}_{k_1}\cdots
\Phi^{j_n}_{k_n}\,, \label{sub}
\end{equation}
where, in the above expression, we have written explicitly the $SU(N)$
indices but neglected the R-symmetry ones. These operators have the
correct orthogonality property when $n$ is comparable to $N$
and therefore are good candidates to describe single particle states.
They belong to a short representation preserving half of the total
supersymmetry, more precisely to a chiral family of $SU(4)$ with
$(0,n,0)$ weight. Note that these operators reproduce the correct
bound for the R-charge, saturated by giant gravitons with $n=N$.

One of the possible approaches to find the form of the dual SYM
operator for generalized giant gravitons consists on constructing the
induced dual map on the SYM theory to an isomorphism on AdS$_5$ and
then apply it on a subdeterminant operator to obtain the desired
generalized dual operator. Once the new CFT dual operator is obtained,
one can check that the resulting properties, like for example
supersymmetry, are the expected ones.

\subsection{Isometries and induced CFT transformations}

To obtain the form of the induced dual map, we project the isometry
into the boundary of AdS$_5$. This boundary can be obtained by considering
very large values of the $X_i$ in the embedding (\ref{AdS}).
Taking $X_i=R\tilde X_i$ in the limit
$R\rightarrow\infty$, this condition becomes
\begin{equation}
\tilde X_0^2-\tilde X_1^2-\tilde X_2^2-\tilde X_3^2-\tilde X_4^2+\tilde X_5^2=0
\end{equation}
and the boundary is given by the projective equivalence classes
$X_i\sim tX_i$, $t\in\R$.
Using this identification, we can rescale the coordinates such that
$X_0^2+X_5^5=1$ and hence the boundary of AdS$_5$ is given by
\begin{equation}
  X_0^2+X_5^5=1=X_1^2+X_2^2+X_3^2+X_4^2\,,
\end{equation}
i.e. it is just $\calS^1\times\calS^3$, with the lorentzian induced
metric.

It is convenient to use the euclidean version of AdS$_5$ to formulate
the AdS/CFT correspondence. To do this, we rotate $X_5\mapsto iX_5$,
mapping AdS$_5$ into the five-dimensional ball $B_5$,
\begin{equation}
  -X_0^2+X_5^2+\sum_i(X_i)^2=-L^2\,,\qquad i=1\ldots4\,.
\end{equation}
To reach the boundary, we take $X_a\rightarrow\infty$ and define
$X_a=t\tilde X_a$. Then, the boundary is given by $-\tilde
X_0^2+\tilde X^5+\sum_i(\tilde X_i)^2=0$ with the identification $\tilde
X\sim\lambda\tilde X$.
The new coordinates $\tilde u=\tilde X_0+\tilde X_5$ and
$\tilde v=\tilde X_0-\tilde X_5$ are such that $\tilde u\tilde
v=\tilde X_i\tilde X_i$, and we can use the projective equivalence
to set $\tilde v=1$. Then $\tilde u=X_iX_i$, and the boundary is
spanned by the coordinates $X_i$, $i=1\ldots4$, endowed with the
euclidean metric.

Let us see how isometry transformations are mapped to the
boundary of euclidean AdS$_5$ in these coordinates. We shall restrict,
for sake of simplicity, to infinitesimal isometries.
The finite transformations can in any case be recovered by exponentiation.
Transformations generated by the hamiltonian (time translations) are
generated by the infinitesimal transformation
\begin{equation}
  X_0'=X_0+\eps X_5,\qquad X_5'=X_5+\eps X_0,\qquad X_i'=X_i\,.
\end{equation}
The induced transformation on the boundary is then
\begin{equation}
  \tilde u'=(1+\eps)\tilde u,
  \qquad\tilde v'=1-\eps,\qquad
  \tilde X_i'=\tilde X_i,
\end{equation}
and we multiply by $(1+\eps)$ to choose the equivalence class representative
with $\tilde v'=1$,
\begin{equation}
  \tilde u'=(1+2\eps)\tilde u,\qquad
  \tilde v'=1,\qquad
  \tilde X_i'=(1+\eps)\tilde X_i.
\end{equation}
Hence, on the boundary, the AdS hamiltonian generates a dilatation
$X_i'=\lambda X_i$, and the energy is mapped in the conformal weight of the
dual operator $\OO$.

Consider now the transformation $R_{(0,3)}(\rho_0)\circ R_{(4,5)}(\rho_0)$
which maps the $\rho=0$ geodesic to the geodesic
$\{\psi=t,\rho=\rho_0\}$. The infinitesimal transformation reads
\begin{eqnarray}
  X_0'=X_0+\eps X_3,&
  X_3'=X_3+\eps X_0,\nonumber\\
  X_4'=X_4+\eps X_5,&
  X_5'=X_5-\eps X_4,\nonumber\\
  X_1'=X_1,&
  X_2'=X_2,
\end{eqnarray}
and acts on the boundary as
\begin{eqnarray}
  u'=u+\eps(X_3-X_4)&
  v'=1+\eps(X_3+X_4)\nonumber\\
  X_3'=X_3+\frac\eps2 (u+v)& \displaystyle
  X_4'=X_4+\frac\eps2 (u-v)\nonumber\\
  X_1'=X_1,&
  X_2'=X_2\,.
\end{eqnarray}
Multiplying by $1-\eps(X_3+X_4)$ to take $v'=1$,
\begin{eqnarray}
  u'=u+\eps(X_3-X_4)-\eps u(X_3+X_4),&
  v'=1\nonumber\\
  X_3'=X_3+\frac\eps2 (u+v)-\eps X_3(X_3+X_4), & \displaystyle
  X_1'=(1-\eps u(X_3+X_4))X_1,\nonumber\\
  X_4'=X_4+\frac\eps2 (u-v)-\eps X_4(X_3+X_4), & \displaystyle
  X_2'=(1-\eps u(X_3+X_4))X_2,\nonumber\\
&
\end{eqnarray}
and using $v=1$, $u=(X_i)^2$, we obtain the action of the transformation
on the boundary,
\begin{eqnarray}
  X_1'=\left[1-\eps (X_3+X_4)\right]X_1,&\nonumber\\
  X_2'=[1-\eps (X_3+X_4)]X_2,&\nonumber\\
  X_3'=[1-\eps (X_3+X_4)]X_3+\frac\eps2 (X^2+1),&\nonumber\\
  X_4'=[1-\eps (X_3+X_4)]X_4+\frac\eps2 (X^2-1).&
\label{isodual}
\end{eqnarray}
Remember that the general infinitesimal $SO(5,1)$ transformation can be
parameterized as
\begin{equation}
  X_i'=(1+\lambda-\beta\cdot X)X_i
  +\frac{X^2}2\beta_i+\frac12\alpha_i+\om_{ij}X_j,
\end{equation}
with $\alpha_i$, $\omega_{ij}$, $\lambda$ and $\beta_i$ the infinitesimal
parameter generating the translations, rotations, dilatations and special
conformal transformations respectively. Hence, we see that the transformation
under consideration is obtained with
\begin{equation}
  \al_i=(0,0,\eps,-\eps),\quad \om_{ij}=0,\quad
  \lambda=0,\quad \beta_i=(0,0,\eps,\eps),
\label{transf}\end{equation}
i.e. it is a combination of a translation with a special conformal
transformation.

\subsection{Generalized giant gravitons and transformed subdeterminants}

Once we know the CFT form of the induced transformation corresponding
to an isometry of AdS$_5$, we just have to consider its action on
the dual operator (\ref{sub}) of a giant graviton sitting on the
center of AdS$_5$ to obtain the form of the dual operator of the
generalized giant graviton. Since this induced transformation is an
element of the conformal group, we know how it acts on any CFT
operator. For example, on field realizations that are eigenfunctions
of the dilatation operator with scaling dimension $\Delta$, like
$\Phi^I(x)$, we have
\begin{equation}
\Phi'^I(x')=\left|\frac{\p x'}{\p x}\right|^{-\Delta/4} \Phi^I(x)\,,
\qquad x'=\Lambda(x)\,,
\end{equation}
where the prime indicates transformed quantities and $\Lambda$ is the
induced conformal map on the four-dimensional space-time coordinates. In
particular, consider the infinitesimal transformation of equation
(\ref{isodual}) for the above scalar fields $\Phi^I(x)$ evaluated at $x=0$,
\begin{equation}
\Phi'^I(0)=(1-\alpha^i\partial_i)\Phi^I(0)\ .
\end{equation}
Due to the fact that subdeterminants are made out of these fields, as
shown in (\ref{sub}), their infinitesimal transformation is just the same.
Note that the transformed operator has the same conformal weight
as the original operator, in agreement with the supergravity picture
where the momentum along the $\calS^5$ directions has not been
modified. The action of the map is just to add a descendent field
part to the original operator, as should be the case for a bosonic
transformation acting on a representation of the superconformal
group. Also, note that in the transformed subdeterminant a space-time
scale $\alpha$ that was absent before appears, signing the fact that
our giant graviton is not any more in the center on AdS$_5$.

The supersymmetries preserved by transformed subdeterminants have to
be the same as the ones of the original subdeterminants because, by
construction, we have just acted with an element of the conformal
group, and therefore we still have operators in the same
supermultiplet. This can be explicitly seen by acting on the
corresponding supersymmetry constraint equation. We know the action of
half of the $Q$ on the subdeterminant should vanish, hence
\begin{equation}
\left[Q,\det{}_n(\Phi)\right]=0 \longrightarrow
U\left[Q,\det{}_n(\Phi)\right]U^{-1}=0\,,
\end{equation}
that in turn implies that
\begin{equation}
\left[Q',\det{}_n(\Phi)'\right]=0\,,
\end{equation}
where we have used the general form of the transformation law
$Q'=UQU^{-1}$, where $U$ is the relevant representation
of the group element $\Lambda$. To be more precise, in
the particular case of the infinitesimal transformation
(\ref{isodual}) acting for example on $Q^I_\alpha$ (where I is
in the $SU(4)$ index and $\al$ is a Weyl-spinor index), we get
\begin{equation}
Q'^I_\alpha=Q^I_\alpha +\beta_i\si_{\al\dot\al}^i\bar S^{I\,\dot \al}\,,
\end{equation}
where $\be_i$ is the generator of the special conformal transformation
(\ref{transf}).
Observe that we have just mixed $Q^I_\al$ and $\bar S^{I\dot\al}$ that
indeed form by themselves a representation of the conformal group
(see for example \cite{Dolan:2002zh}). Therefore, we can safely
conclude that generalized giant gravitons correspond to half
BPS states in the CFT.

Next, we move on into the question of the dual operators to
multiparticle giant graviton states breaking one quarter of the
supersymmetries proposed in section \ref{quarter}.
As dual CFT operators to these composite giant graviton states,
we propose the naive product of two subdeterminants, where the second
one has been transformed by the map induced by the AdS$_5$ isometry,
i.e.
\begin{equation}
\OO_{(n,k)}=\det{}_n(\Phi)\times\det{}_k(\Phi)'\,,
\label{quarterCFT}\end{equation}
where $n$ and $k$ are the R-charges of the first and second giant
gravitons respectively. Note that this candidate operator has
the correct R-charge $n+k$, and encodes the characteristic bound for
each giant graviton.
The R-symmetry properties of these composite states are those of the
tensor product of the representations $(0,n,0)$ and $(0,k,0)$ in which
lie the two giant gravitons. As shown in appendix \ref{app}, the
representations appearing in the above product are of the form
$(p,q,p)$, some of which are short and define one quarter BPS states 
\cite{Andrianopoli}, while the others are long representations
not enjoying any protection from quantum corrections.
The fact that the supergravity binary system is supersymmetric
implies, from the AdS/CFT correspondence, that also the dual CFT
operator should be supersymmetric. Hence, long representations should
not occur in the above decomposition, and the resulting dual operator
should be in a {\em short} $(p,q,p)$ representation, preserving therefore
exactly one quarter of the supersymmetry. This fraction coincides with
the one found in the supergravity description of section
\ref{quarter}.
Although the above argument is not a complete proof, it is certainly a
check that the proposed dual operator have passed.

Note that group theory arguments imply the supersymmetry of these
operators only for the free theory. In the interacting case, quarter
BPS operators are known to mix with other descendent non-BPS
operators. Nevertheless, in our case the duality with the supergravity
picture suggests that the operator $\OO_{(n,k)}$ is one quarter BPS
for the full interactive theory in the infinite $N$ and infinite
't~Hooft parameter limit.

As a last comment, note that our quarter BPS operators are made out of
subdeterminants and therefore are valid for large R-charge, while
those explicitly considered in the literature have small R-charge
\cite{Andrianopoli,Ryzhov}.


\section{Summary and discussion}

In this article, we have found giant graviton configurations with generic
motion in AdS$_5$.
The D3-brane dynamics on the giant graviton embedding ansatz reduces
to that of a massive point particle in AdS$_5$ and therefore all the
corresponding solutions follow timelike geodesics. Due to the fact that
all such geodesics are related via isomorphism transformations of the
background fields, the most general giant graviton configuration can be
found by acting on the original giant graviton solution of
ref.~\cite{McGreevy:2000cw}. In particular, to illustrate better this
solution-generating mechanism, we considered the explicit example of a
giant graviton orbiting on a great circle of the $\calS^3$ inside
AdS$_5$ at a constant radius.
Nevertheless, we emphasize the fact that any timelike geodesic
solves the problem and therefore more involved configurations, having
multiple angular momenta and oscillating radial position, are possible.

Next, we proved that all these new solutions are one half BPS states,
by a detailed analysis on the meaning of an isometry transformation
for the $\kappa$-symmetry constraint of the D3-brane supersymmetric
world volume theory. Again, for the case of the orbiting giant
graviton, the explicit form of the $\kappa$-symmetry projector and of
the surviving Killing spinor is given.

An interesting outcome of the above analysis is the observation that
different giant gravitons, following different timelike geodesics,
have different $\kappa$-symmetry projectors. It is therefore possible
to break larger fractions of supersymmetry by considering
multiparticle giant graviton states. To illustrate this mechanism, we
explicitly constructed a quarter BPS binary system, with one giant
graviton orbiting around the other. We would like to stress that this
configuration implements a new kind of time-dependent supersymmetric
solution in string theory, since time-dependent states will arise
in the open string sector for strings stretched between the two giant
gravitons (see \cite{Myers:2002bk} for a similar construction).

Also, using the AdS/CFT conjecture, we obtained the explicit form of
the dual operator to generalized giant gravitons, and showed it is half
BPS. In particular, this operator can be seen as the result of a conformal
symmetry transformation on subdeterminant operators. Finally, we
proposed the product of such operators to describe multiparticle giant
graviton states in the CFT, and gave arguments in favor of this
hypothesis.

It would be interesting to extend the present work to verify, by a
direct calculation, that these multiparticle operators indeed
preserve one quarter of the supersymmetries. We believe that, using
these techniques, it should be possible to obtain other fractions of
supersymmetry, like for example one eighth.
As a matter of fact, one eighth BPS states can be singled out from
products of three or more single trace half BPS operators\footnote{See
  section 3.5 of \cite{D'Hoker:2002aw} for a nice review, and
  references therein.}.
Since in the low momentum limit the CFT dual of giant gravitons should
reduce to single trace operators, it is tempting to conjecture that
configurations of three giant gravitons preserving one eighth of the
total supersymmetry could be found.
The corresponding supergravity problem is to find a three giant
graviton configuration preserving one eighth of the supersymmetries.

Finally, it is well-known that standard giant graviton condensates
give rise to superstars \cite{superstars}. In this article we did not
consider the backreaction on the geometry due to the presence of the
D-branes.
Nevertheless, it an important subject; since generalized giant
gravitons carry angular momentum in AdS$_5$, some of their condensates
may form a rotating generalization of the superstar.
Work is in progress to identify such a supergravity solution.


\section*{Acknowledgements}
\small
The authors would like to thank D.~Klemm, L.~Martucci, and A.~Santambrogio
for useful discussions.

This work was partially supported by INFN, MURST and by the European
Commission RTN program HPRN-CT-2000-00131, in which M.~M.~C. and
P.~J.~S. are associated to the University of Torino.
\normalsize


\appendix

\section{Some properties of product states}
\label{app}

\noindent{\bf Theorem 1:}
{\sl In the decomposition of the tensor product $(0,p,0)\otimes(0,q,0)$ into
  irreducible representations, only representations of the form $(m,n,m)$
  occur.}\\
\noindent{\bf Proof:}
A $su(4)$ highest weight with Dynkin labels $(0,p,0)$ can equally well be
specified in terms of its partition $\{p;p\}$ corresponding to a Young tableau
with two rows and $p$ boxes in each row.
To compute the tensor product $(0,p,0)\otimes(0,q,0)$ we can then apply the
Littlewood-Richardson rule to the product of Young tableaux
$$
\underbrace{
  \btexdraw
  \drawdim cm
  \move(0 -.5)\lvec(3 -.5)\lvec(3 .5)\lvec(0 .5)\lvec(0 -.5)
  \move(0 0)\lvec(3 0)
  \move( .5 -.5)\lvec(.5  .5)
  \move(1   -.5)\lvec(1   .5)
  \move(1.5 -.5)\lvec(1.5 .5)
  \move(2.5 -.5)\lvec(2.5 .5)
  \move(1.8 .25)\fcir f:0 r:.01
  \move(2.0 .25)\fcir f:0 r:.01
  \move(2.2 .25)\fcir f:0 r:.01
  \move(1.8 -.25)\fcir f:0 r:.01
  \move(2.0 -.25)\fcir f:0 r:.01
  \move(2.2 -.25)\fcir f:0 r:.01
  \move(0 -0.6)
  \etexdraw
}_{p}
\quad\raisebox{5mm}{$\otimes$}\quad
\underbrace{
  \btexdraw{
    \drawdim cm
    \move(0 -.5)\lvec(3 -.5)\lvec(3 .5)\lvec(0 .5)\lvec(0 -.5)
    \move(0 0)\lvec(3 0)
    \move( .5 -.5)\lvec(.5  .5)
    \move(1   -.5)\lvec(1   .5)
    \move(1.5 -.5)\lvec(1.5 .5)
    \move(2.5 -.5)\lvec(2.5 .5)
    \move(1.8 .25)\fcir f:0 r:.01
    \move(2.0 .25)\fcir f:0 r:.01
    \move(2.2 .25)\fcir f:0 r:.01
    \move(1.8 -.25)\fcir f:0 r:.01
    \move(2.0 -.25)\fcir f:0 r:.01
    \move(2.2 -.25)\fcir f:0 r:.01
    \textref h:C v:C
    \htext(.25 .25){a}
    \htext(.75 .25){a}
    \htext(1.25 .25){a}
    \htext(2.75 .25){a}
    \htext(.25 -.25){b}
    \htext(.75 -.25){b}
    \htext(1.25 -.25){b}
    \htext(2.75 -.25){b}
    \move(0 -0.6)
  } \etexdraw
}_{q}
$$
We first have to add the $q$ boxes \ybox{a} of the right tableau to the
left tableau; we add $i$ boxes to the first row, and $q-i$ to the third row. No
box can be added to the second row because we cannot have boxes with the same
label in the same column. We obtain a Young tableau with partition
$\{p+i;p;q-i\}$. Regularity imposes then $p\geq q-i$.
We have now to add $q$ boxes \ybox{b} to the resulting tableau.
In counting from right to left and top to bottom, the number of boxes
\ybox{a} must always be greater or equal to the number of boxes
\ybox{b}. Therefore, no \ybox{b} box can be added to the first line;
let us call $j$ and $k$ the number of \ybox{b} boxes added to the
second and third line respectively. The remaining $q-j-k$ \ybox{b}
boxes are inserted in the fourth line. We end with a Young tableau
with partition $\{p+i;p+j;q-i+k;q-j-k\}$. Now, to keep the counting of
\ybox{a} greater than the number of \ybox{b}, we must impose $j\leq i$
and $j+k\leq i$; the request that no two \ybox{b} boxes fall into the
same column requires on the other hand $q-i+k\leq p$ and $q-j-k\leq q-i$.
Finally, regularity of the tableau imposes $p+i\geq p+j\geq q-i+k\geq q-j-k$.
Now, combining these inequalities, it follows that $i=j+k$, and the partition
reads $\{p+j+k;p+j;q-j;q-j-k\}$. The first $q-j-k$ columns of the tableau have
four rows, and can be ignored by eliminating $q-j-k$ boxes to each row
(these representations are the long ones). Hence, the most general
Young tableau appearing in the tensor product has partition
$\{p-q+2j+2k;p-q+2j+k;k\}$, or equivalently, Dynkin labels
$(k,p-q+2j,k)$, of the form $(m,n,m)$ as stated.
\phantom{.}\hfill$\Box$

Note that short $(0,q,0)$ states have scaling dimension $\Delta=q$,
and short $(p,q,p)$ states have scaling dimension $\Delta=2p+q$.
The short representation $(k,p-q+2j,k)$ has conformal scaling dimension
$\Delta=p-q+2j+2k$, and requiring it to be 
equal to the conformal dimension $p+q$ of the product $(0,p,0)\otimes(0,q,0)$,
we obtain the relation $2j+2k=p+2q$. In other words, the only short $SU(4)$
multiplets appearing in the tensor product are of the form $(k,2p+q-2k,k)$ with
$k$ positive integer subject to the constraints $k\leq q$ and $2k\leq3p$.



\begin{thebibliography}{99}


\bibitem{Petersen:1999zh}
J.~L.~Petersen,
``Introduction to the Maldacena conjecture on AdS/CFT,''
Int.\ J.\ Mod.\ Phys.\ A {\bf 14} (1999) 3597
[arXiv:hep-th/9902131].

\bibitem{Aharony:1999ti}
O.~Aharony, S.~S.~Gubser, J.~M.~Maldacena, H.~Ooguri and Y.~Oz,
``Large N field theories, string theory and gravity,''
Phys.\ Rept.\  {\bf 323} (2000) 183
[arXiv:hep-th/9905111].

\bibitem{D'Hoker:2002aw}
E.~D'Hoker and D.~Z.~Freedman,
``Supersymmetric gauge theories and the AdS/CFT correspondence,''
arXiv:hep-th/0201253.

\bibitem{McGreevy:2000cw}
J.~McGreevy, L.~Susskind and N.~Toumbas,
``Invasion of the giant gravitons from anti-de Sitter space,''
JHEP {\bf 0006} (2000) 008
[arXiv:hep-th/0003075].

\bibitem{Grisaru:2000zn}
M.~T.~Grisaru, R.~C.~Myers and O.~Tafjord,
``SUSY and Goliath,''
JHEP {\bf 0008} (2000) 040
[arXiv:hep-th/0008015].

\bibitem{Myers:1999ps}
R.~C.~Myers,
``Dielectric-branes,''
JHEP {\bf 9912} (1999) 022
[arXiv:hep-th/9910053].

\bibitem{Maldacena:1998bw}
J.~M.~Maldacena and A.~Strominger,
``AdS(3) black holes and a stringy exclusion principle,''
JHEP {\bf 9812} (1998) 005
[arXiv:hep-th/9804085].

\bibitem{dualgg}
A.~Hashimoto, S.~Hirano and N.~Itzhaki,
``Large branes in AdS and their field theory dual,''
JHEP {\bf 0008} (2000) 051
[arXiv:hep-th/0008016];\\
S.~Corley, A.~Jevicki and S.~Ramgoolam,
``Exact correlators of giant gravitons from dual N = 4 SYM theory,''
Adv.\ Theor.\ Math.\ Phys.\  {\bf 5} (2002) 809
[arXiv:hep-th/0111222].

\bibitem{superstars}
R.~C.~Myers and O.~Tafjord,
``Superstars and giant gravitons,''
JHEP {\bf 0111} (2001) 009
[arXiv:hep-th/0109127];\\
F.~Leblond, R.~C.~Myers and D.~C.~Page,
``Superstars and giant gravitons in M-theory,''
JHEP {\bf 0201} (2002) 026
[arXiv:hep-th/0111178].

\bibitem{Balasubramanian:2001nh}
V.~Balasubramanian, M.~Berkooz, A.~Naqvi and M.~J.~Strassler,
``Giant gravitons in conformal field theory,''
JHEP {\bf 0204} (2002) 034
[arXiv:hep-th/0107119].

\bibitem{Arapoglu:2003ti}
S.~Arapoglu, N.~S.~Deger, A.~Kaya, E.~Sezgin and P.~Sundell,
``Multi-spin giants,''
arXiv:hep-th/0312191.

\bibitem{Bozhilov:2002sj}
P.~Bozhilov,
``Probe branes dynamics: Exact solutions in general backgrounds,''
Nucl.\ Phys.\ B {\bf 656} (2003) 199
[arXiv:hep-th/0211181].

\bibitem{Gibbons:1999uv}
G.~W.~Gibbons and C.~A.~R.~Herdeiro,
``Supersymmetric rotating black holes and causality violation,''
Class.\ Quant.\ Grav.\  {\bf 16} (1999) 3619
[arXiv:hep-th/9906098].

\bibitem{Dorn:2003ct}
H.~Dorn and C.~Sieg,
``Conformal boundary and geodesics for AdS(5) x S**5 and the plane  wave:
Their approach in the Penrose limit,''
JHEP {\bf 0304} (2003) 030
[arXiv:hep-th/0302153].

\bibitem{oneill}
B.~O'Neill, ``Semi-Riemannian Geometry, with Applications to Relativity,''
Academic Press, San Diego, USA,  1983.

\bibitem{Matschull:1998rv}
H.~J.~Matschull,
``Black hole creation in 2+1-dimensions,''
Class.\ Quant.\ Grav.\  {\bf 16} (1999) 1069
[arXiv:gr-qc/9809087].

\bibitem{M2}
E.~Bergshoeff, E.~Sezgin and P.~K.~Townsend,
``Supermembranes And Eleven-Dimensional Supergravity,''
Phys.\ Lett.\ B {\bf 189} (1987) 75;
``Properties Of The Eleven-Dimensional Super Membrane Theory,''
Annals Phys.\  {\bf 185} (1988) 330.

\bibitem{M5}
I.~A.~Bandos, K.~Lechner, A.~Nurmagambetov, P.~Pasti, D.~P.~Sorokin and M.~Tonin,
``Covariant action for the super-five-brane of M-theory,''
Phys.\ Rev.\ Lett.\  {\bf 78} (1997) 4332
[arXiv:hep-th/9701149];
``On the equivalence of different formulations of the M theory  five-brane,''
Phys.\ Lett.\ B {\bf 408} (1997) 135
[arXiv:hep-th/9703127];\\
M.~Aganagic, J.~Park, C.~Popescu and J.~H.~Schwarz,
``World-volume action of the M-theory five-brane,''
Nucl.\ Phys.\ B {\bf 496} (1997) 191
[arXiv:hep-th/9701166];\\
P.~S.~Howe and E.~Sezgin,
``Superbranes,''
Phys.\ Lett.\ B {\bf 390} (1997) 133
[arXiv:hep-th/9607227];
``D = 11, p = 5,''
Phys.\ Lett.\ B {\bf 394} (1997) 62
[arXiv:hep-th/9611008];\\
P.~S.~Howe, E.~Sezgin and P.~C.~West,
``Covariant field equations of the M-theory five-brane,''
Phys.\ Lett.\ B {\bf 399} (1997) 49
[arXiv:hep-th/9702008].

\bibitem{D}
M.~Cederwall, A.~von Gussich, B.~E.~W.~Nilsson and A.~Westerberg,
``The Dirichlet super-three-brane in ten-dimensional type IIB  supergravity,''
Nucl.\ Phys.\ B {\bf 490} (1997) 163
[arXiv:hep-th/9610148];\\
M.~Cederwall, A.~von Gussich, B.~E.~W.~Nilsson, P.~Sundell and A.~Westerberg,
``The Dirichlet super-p-branes in ten-dimensional type IIA and IIB
supergravity,''
Nucl.\ Phys.\ B {\bf 490} (1997) 179
[arXiv:hep-th/9611159];\\
E.~Bergshoeff and P.~K.~Townsend,
``Super D-branes,''
Nucl.\ Phys.\ B {\bf 490} (1997) 145
[arXiv:hep-th/9611173].

\bibitem{Ferrara:1999ed}
S.~Ferrara and A.~Zaffaroni,
``Superconformal field theories, multiplet shortening, and the  AdS(5)/SCFT(4)
correspondence,''
arXiv:hep-th/9908163.

\bibitem{Dolan:2002zh}
F.~A.~Dolan and H.~Osborn,
``On short and semi-short representations for four dimensional superconformal
symmetry,''
Annals Phys.\  {\bf 307} (2003) 41
[arXiv:hep-th/0209056].

\bibitem{Andrianopoli}
L.~Andrianopoli and S.~Ferrara,
``Short and long SU(2,2/4) multiplets in the AdS/CFT correspondence,''
Lett.\ Math.\ Phys.\  {\bf 48} (1999) 145
[arXiv:hep-th/9812067];\\
L.~Andrianopoli, S.~Ferrara, E.~Sokatchev and B.~Zupnik,
``Shortening of primary operators in N-extended SCFT(4) and
harmonic-superspace analyticity,''
Adv.\ Theor.\ Math.\ Phys.\  {\bf 3} (1999) 1149
[arXiv:hep-th/9912007].

\bibitem{Skiba:1999im}
W.~Skiba,
``Correlators of short multi-trace operators in N = 4 supersymmetric
Yang-Mills,''
Phys.\ Rev.\ D {\bf 60} (1999) 105038
[arXiv:hep-th/9907088].

\bibitem{Ryzhov}
A.~V.~Ryzhov,
``Quarter BPS operators in N = 4 SYM,''
JHEP {\bf 0111} (2001) 046
[arXiv:hep-th/0109064];
E.~D'Hoker, P.~Heslop, P.~Howe and A.~V.~Ryzhov,
``Systematics of quarter BPS operators in N = 4 SYM,''
JHEP {\bf 0304} (2003) 038
[arXiv:hep-th/0301104].

\bibitem{Myers:2002bk}
R.~C.~Myers and D.~J.~Winters,
``From D - anti-D pairs to branes in motion,''
JHEP {\bf 0212} (2002) 061
[arXiv:hep-th/0211042].

\end{thebibliography}
\end{document}